\documentclass[10pt, doublecolumn]{IEEEtran}
\usepackage{epsfig,latexsym}
\usepackage{float}
\usepackage{indentfirst}
\usepackage{amsmath}
\usepackage{bm}
\usepackage{amssymb}
\usepackage{times}
\usepackage{enumitem}

\usepackage{algorithm}
\usepackage[noend]{algpseudocode}
\usepackage{subfigure}
\usepackage{psfrag}
\usepackage{hyperref}
\usepackage{cite}
\usepackage{lastpage}
\usepackage{fancyhdr}
\usepackage{color} 
 \usepackage{amsthm}
\usepackage{bigints}
\usepackage{array}
\usepackage{booktabs}
\newtheorem{Lemma}{Lemma}
\newtheorem{Corollary}{Corollary}
\newtheorem{lemma}[Lemma]{$\mathbf{Lemma}$}

\newtheorem{corollary}[Corollary]{$\mathbf{Corollary}$}

\newcounter{problem}
\newcounter{save@equation}
\newcounter{save@problem}
\makeatletter

\begin{document}
\title{  \vspace{-1em}\huge{  LoS Blockage in Pinching-Antenna Systems: \\Curse or Blessing? }}

\author{ Zhiguo Ding, \IEEEmembership{Fellow, IEEE}, and H. Vincent Poor, \IEEEmembership{Life Fellow, IEEE}   \thanks{ 
  
\vspace{-2em}

Z. Ding is with the University
of Manchester, Manchester, M1 9BB, UK, and Khalifa University, Abu Dhabi, UAE.    
H. V. Poor is  with the  Department of Electrical and Computer Engineering, Princeton University,
Princeton, NJ 08544, USA.
 

  }\vspace{-2.6em}}
 \maketitle

\begin{abstract}
This letter investigates the impact of line-of-sight (LoS) blockage on pinching-antenna systems.  Analytical results are developed for both single-user and multi-user cases to reveal that the presence of LoS blockage is beneficial for increasing the performance gain of pinching antennas over conventional antennas. This letter also reveals that LoS blockage is particularly useful in multi-user cases, where co-channel interference can be effectively suppressed by LoS blockage.   
\end{abstract}\vspace{-0.2em}

\vspace{-1.3em} 

\section{Introduction}
Pinching-antenna systems have recently been recognized as a promising transmission technique for next-generation mobile networks due to the following three features \cite{pinching_antenna2,pamagazine}. The first feature is their low costs since pinching antennas are simple dielectric particles, e.g., clothes pinches, applied on waveguides. The second feature is the capability of pinching antennas to create strong line-of-sight (LoS) connections between a base station and its user, e.g., it is possible to activate a pinching antenna right next to the user, and hence the path loss experienced by the user can be very small. The third feature is that the multi-input multi-output (MIMO) systems created by pinching antennas can be flexibly reconfigured, e.g., adding (or removing) antennas becomes straightforward.  

The fundamental limits of pinching-antenna systems with different configurations, e.g., different numbers of pinching antennas, users, and waveguides, have been identified in \cite{mypa}. These obtained analytical results reveal that pinching antennas achieve a significant performance gain over conventional antennas. In \cite{kaidpa}, a low-complexity implementation to activate pinching antennas, instead of moving them, was studied, and the array gain achieved by pinching-antenna systems has been identified in \cite{Chongjunpa1}. In addition, sophisticated resource allocation algorithms have been developed for uplink and downlink transmissions in pinching-antenna systems \cite{yanqingpa} and \cite{Tegosuppa}, respectively. 

Recall that one of the key features of pinching antennas is to reduce the transceiver distance, which means that in pinching-antenna systems, a user experiences less large-scale path loss and LoS blockage, compared to conventional antenna systems. However, in the literature, there is no study that formally investigated the impact of pinching antennas on LoS blockage, which motivates this letter.  The contribution of this letter is two-fold. One is to focus on the single-user special case, where analytical results are developed to analyze the outage probability achieved by pinching antennas. The presented analytical and simulation results confirm the intuition that the presence of LoS blockage is beneficial for increasing the performance gain of pinching antennas over conventional antennas, compared to the case without blockage. The second contribution focuses on a general multi-user scenario. For conventional antenna systems, the existence of strong co-channel interference severely degrades the system performance. The presence of LoS blockage makes it more difficult to combat co-channel interference since the LoS blockage can make a user's channel matrix no longer full rank, and hence, many interference cancellation methods, such as zero-forcing approaches, become not applicable. However, in pinching-antenna systems, LoS blockage becomes a blessing since a user's interference link is likely to be subject to blockage since the corresponding interfering pinching antenna could be far away from the user. This intuition is confirmed by the presented analytical results, which show that the ergodic data rate gain of pinching antennas over conventional antennas is unbounded at a high signal-to-noise ratio (SNR).
\vspace{-1em} 
\section{System Model}\label{section 2}
 Consider a downlink pinching-antenna system with $M$ single-antenna users, denoted by ${\rm U}_m$, in a rectangular-shaped service area, denoted by $\mathcal{A}$, whose two sides are denoted by $D_{\rm W}$ and $D_{\rm L}$. Assume that the service area is divided by $M$ parallelly installed waveguides into $M$ identical rectangles with sides being $\frac{D_{\rm W}}{M}$ and $D_{\rm L}$. To facilitate the performance analysis, assume that ${\rm U}_m$ is uniformly distributed in the rectangle centered by the $m$-th waveguide, and ${\rm U}_m$'s location is denoted by $\boldsymbol{\psi}_m= (x_m,y_m,0)$. It is further assumed that a single pinching antenna is activated on the $m$-th waveguide at the location closest to ${\rm U}_m$, and hence its location can be denoted by  ${\boldsymbol \psi}_m^{\rm Pin}=(x_m,\beta_m,d)$, where $d$ denotes the height of the waveguide and $\beta_m=-\frac{D_{\rm W}}{2}+(m-1)\frac{D_{\rm W}}{M}+\frac{D_{\rm W}}{2M}$.   
 
Similar to \cite{mypa}, ${\rm U}_m$'s observation is given by 
{\small  \begin{align}\label{models}
  y_m =& \sum^{M}_{k=1}\tilde{h}_{mk} p_{mk} \sqrt{P}s_m +\sum_{i\neq m}\sum^{M}_{k=1}\tilde{h}_{mk}  p_{ik} \sqrt{P}s_i +w_m,
  \end{align}}
  \hspace{-0.9em}where $\tilde{h}_{mk}=\alpha_{mk}h_{mk}$, $h_{mk} $ denotes the channel gain between the $k$-th antenna and ${\rm U}_m$, i.e., $h_{mk}=\frac{\sqrt{\eta} e^{-2\pi j \left(\frac{  1}{\lambda}\left| {\boldsymbol \psi}_m  - {\boldsymbol \psi}_k^{\rm Pin}\right|
  +\frac{1}{\lambda_g}\left| {\boldsymbol \psi}_0^{\rm Pin}  - {\boldsymbol \psi}_k^{\rm Pin}\right|
  \right)}}{  \left| {\boldsymbol \psi} _m - {\boldsymbol \psi}_k^{\rm Pin}\right|} $, the overall transmit power budget is denoted by $P$,  $p_{mk}$ is the precoding coefficient,  $\lambda$ and $\lambda_g$ denote the carrier and waveguide wavelengths, respectively, $\eta = \frac{c^2}{16\pi^2 f_c^2 }$, $c$ is the speed of light, the carrier frequency is denoted by $f_c$, $w_m$ denotes the additive noise with power $\sigma^2$,  $s_m$ denotes ${\rm U}_m$'s signal, and $\alpha_{mk}$ is an indicator function for the LoS blockage. In particular, if there is LoS blockage between the pinching antenna on the $k$-th waveguide and ${\rm U}_{m}$, $\alpha_{mk}=0$. Otherwise, $\alpha_{mk}=1$. In general, LoS blockage can be modeled as follows: \cite{5621983}
\begin{align}\label{blockage}
 \mathbb{P}(\alpha_{mk}=1) = e^{-\phi |\boldsymbol{\psi}^{\rm Pin}_i - \boldsymbol{\psi}_m| } ,
\end{align}
and for ultra-dense indoor environments, the following LoS blockage model can also be used: \cite{3gppblock}
\begin{align}\label{blockage2}
 \mathbb{P}(\alpha_{mk}=1) = e^{-\phi |\boldsymbol{\psi}^{\rm Pin}_i - \boldsymbol{\psi}_m|^2 } ,
\end{align}
where $\phi$ denotes the LoS blockage parameter and is decided by the communication environment.  We note that the waveguide propagation loss is omitted in \eqref{models}, but will be considered in the simulation section. There are two possible system designs, as described in the following subsections.

\vspace{-1em}
 \subsection{System Design I}
Zero-forcing-based precoding can be applied, where  $ y_m$ needs to be expressed in a more compact matrix form:
  \begin{align}\label{matrix model}
  y_m =& \mathbf{h}_m^H\mathbf{P}\mathbf{s}+w_m,
  \end{align}
  where $\mathbf{h}_m=\begin{bmatrix} \tilde{h}_{m1}&\cdots &\tilde{h}_{mM}\end{bmatrix}^H$, the element at the $k$-th row and the $m$-th column of $\mathbf{P}$ is $p_{mk}$, and $\mathbf{s}$ collects all the signals.  
By assuming that $\mathbf{P}$ is full-rank, the zero forcing precoding can be applied, which means that $\mathbf{P}=\mathbf{H}^{-H}\mathbf{D}$, where the $m$-th column of $\mathbf{H}$ is $\mathbf{h}_m$, and $\mathbf{D}$ is an $M\times M$ diagonal matrix to ensure the power normalization, i.e., $[\mathbf{D}]_{m,m}=\sqrt{g_m}$, where $ g_m\triangleq  \frac{1}{M\left[\mathbf{H}^{-1}\mathbf{H}^{-H}\right]_{m,m} }$, and $[\mathbf{A}]_{m,m}$ denotes the $m$-th diagonal element of $\mathbf{A}$. 

Therefore, if $\mathbf{P}$ is full-rank, the effective channel gain experienced by ${\rm U}_m$ is given by $g_m$, which means that  ${\rm U}_m$'s data rate is simply the following: $
  R^{\rm Pin}_m = \log_2\left(
1+ \frac{g_m P}{\sigma^2}
\right)$.
If $\mathbf{P}$ is rank-deficient due to LoS blockage, the low-complexity design in the following subsection will be used, i.e., Design I is a hybrid scheme between zero forcing and Design II.

\vspace{-1em}
\subsection{System Design II}
A low-complexity design is to ask each pinching antenna to serve a single user, i.e., $p_{mm}=\frac{1}{\sqrt{M}}$, and $p_{mk}=0$, for $k\neq m$. With this system design, ${\rm U}_m$'s data rate expression can be simplified as follows: 
\begin{align}\label{1pa1wgdfdd}
  R^{\rm Pin}_m =& \log_2\left(
1+ \frac{ \alpha_{mm}\left|h_{mm} \right|^2P}
{\sum_{i\neq m}  \alpha_{mi}\left|
h_{mi}   
\right|^2P+M\sigma^2}
\right).
\end{align}
 Compared to Design I, the precoding of Design II can be devised with low complexity, as it avoids channel matrix inversion. In the presence of non-line-of-sight (NLoS) paths or small-scale fading, Design II can further reduce the system complexity since the base station needs to know the users' locations only, whereas Design I needs sophisticated channel estimation in order to acquire perfect channel state information at the base station.

\section{Performance Analysis for the  Case of $M=1$}\label{section 3}
Assume that there is a single user and a single waveguide, which means that the user's data rate is simplified as follows:
 \begin{align}\label{1pa1wg3}
  R^{\rm Pin}= \log_2\left(
1+  \frac{\alpha \eta P}{ \sigma^2|\boldsymbol{\psi}^{\rm Pin} - \boldsymbol{\psi}| ^{ 2 } } 
\right),
\end{align}
where the subscription $m$ is omitted. For this special case, the outage probability is adopted as the performance metric, since it can clearly reveal the impact of LoS blockage on the system performance. In particular, the corresponding outage probability can be expressed as follows:
\begin{align}
 \mathbb{P}^{\rm out} =   \mathbb{P}\left(\log_2\left(
1+  \frac{\alpha \eta P}{ \sigma^2|\boldsymbol{\psi}^{\rm Pin} - \boldsymbol{\psi}| ^{ 2 } } 
\right)\leq R^{\rm target}\right),
\end{align}
where $ R^{\rm target}$ denotes the target data rate. 
Since $\alpha$ is a discrete random variable with two possible values, the outage probability can be first expressed as follows: 
\begin{align}
 \mathbb{P}^{\rm out} =&   \mathbb{P}\left(\alpha=0\right)+ \mathbb{P}\left( 
  \frac{ \eta P}{ \sigma^2|\boldsymbol{\psi}^{\rm Pin} - \boldsymbol{\psi}| ^{ 2 } } 
 \leq \epsilon,\alpha=1\right)\\\nonumber
 =&  \mathbb{P}\left(\alpha=0\right)+ \mathbb{P}\left( 
 |\boldsymbol{\psi}^{\rm Pin} - \boldsymbol{\psi}|   
 \geq \tau_1,\alpha=1\right),
\end{align}
where $\epsilon=2^{R^{\rm target}}-1$ and $\tau_1=\sqrt{\frac{\eta P}{\epsilon\sigma^2}}$. 
Based on the commonly used blockage model in \eqref{blockage}, the outage probability can be evaluated as follows:
 \begin{align}
 \mathbb{P}^{\rm out} = & \int_{\boldsymbol{\psi}\in \mathcal{A}} \left(1-e^{-\phi |\boldsymbol{\psi}^{\rm Pin} - \boldsymbol{\psi}| }\right) d \psi  \\\nonumber &+ \int_{\boldsymbol{\psi}\in \mathcal{A},  |\boldsymbol{\psi}^{\rm Pin} - \boldsymbol{\psi}|   
 \geq \tau_1 } e^{-\phi |\boldsymbol{\psi}^{\rm Pin} - \boldsymbol{\psi}| }  d \psi   .
\end{align}

For the single-user special case, the coordinates of ${\rm U}_m$ and the pinching antenna can be simplified as $\boldsymbol{\psi}=(x,y,0)$, and $\boldsymbol{\psi}^{\rm Pin}=(x, 0, d)$, respectively. By applying the assumption that the user is uniformly distributed in the service area $\mathcal{A}$, the outage probability can be expressed as follows:
 \begin{align}\label{all terms}
& \mathbb{P}^{\rm out} =  \frac{1}{D_{\rm W}}\int^{\frac{D_{\rm W}}{2}}_{-\frac{D_{\rm W}}{2}}\left(1-e^{-\phi  \sqrt{y^2+d^2} }\right) d y  \\\nonumber &+  \frac{1}{D_{\rm W}}\int^{\tau_2}_{-\frac{D_{\rm W}}{2}}   e^{-\phi  \sqrt{y^2+d^2}  }  d y   +  \frac{1}{D_{\rm W}}\int_{\tau_3}^{\frac{D_{\rm W}}{2}}   e^{-\phi  \sqrt{y^2+d^2}  }  d y  \\\nonumber
&=1-f\left(
-\frac{D_{\rm W}}{2}, \frac{D_{\rm W}}{2}
\right) +f\left(
-\frac{D_{\rm W}}{2}, \tau_2
\right) +f\left(\tau_3,
\frac{D_{\rm W}}{2} 
\right) ,
\end{align}
where $f\left(
a,b
\right) =  \frac{1}{D_{\rm W}} \int^{b}_{a}   e^{-\phi  \sqrt{y^2+d^2}  }  d y  $, $\tau_2 = \max\left\{
-\frac{D_{\rm W}}{2}, -\sqrt{\tau_1^2-d^2}
\right\}$,
 $\tau_3 = \min\left\{
\frac{D_{\rm W}}{2}, \sqrt{\tau_1^2-d^2}
\right\}$, and the ranges of the second and third integrals are due to the fact that
 $\sqrt{y^2+d^2}\geq\tau_1$ is equivalent to $y\geq \sqrt{\tau_1^2-d^2}$ or $y\leq -\sqrt{\tau_1^2-d^2}$.

At high SNR, $\tau_2\rightarrow -\frac{D_{\rm W}}{w}$ and  $\tau_3\rightarrow \frac{D_{\rm W}}{w}$, and hence the first term in \eqref{all terms} becomes dominant, which yields the following lemma.

\begin{lemma}\label{lemma1}
 For the single-user special case with the blockage model in \eqref{blockage}, the outage probability achieved by the pinching-antenna system can be approximated at high SNR as follows:
 \begin{align}\label{approx}
& \mathbb{P}^{\rm out}  \approx 1-f\left(
-\frac{D_{\rm W}}{2}, \frac{D_{\rm W}}{2}
\right) .
\end{align}
\end{lemma}
%

Lemma \ref{lemma1} can be used to obtain insight into the performance comparison between pinching antennas and conventional antennas. In particular, recall that the outage probability achieved by a conventional antenna with its location fixed at $\boldsymbol{\psi}_0 =(0,0,0)$ is given by 
\begin{align}
 \mathbb{P}^{\rm out}_{\rm conv} =   \mathbb{P}\left(\log_2\left(
1+  \frac{\alpha \eta P}{ \sigma^2|\boldsymbol{\psi}_0- \boldsymbol{\psi}| ^{ 2 } } 
\right)\leq R^{\rm target}\right).
\end{align}

Following steps similar to those for the pinching-antenna case and with some straightforward algebraic manipulations, the outage probability  $ \mathbb{P}^{\rm out}_{\rm conv} $ can be approximated at high SNR as follows:
 \begin{align}\label{conv}
& \mathbb{P}^{\rm out}_{\rm conv} \approx \int^{\frac{D_{\rm L}}{2}}_{-\frac{D_{\rm L}}{2}}\int^{\frac{D_{\rm W}}{2}}_{-\frac{D_{\rm W}}{2}}\left(1-e^{-\phi  \sqrt{x^2+y^2+d^2} }\right) \frac{d ydx   }{D_{\rm W}D_{\rm L}} .
\end{align}

Therefore, the performance gain of pinching-antenna systems over the conventional-antenna system can be expressed as follows:
 \begin{align}\label{conv}
  \mathbb{P}^{\rm out}_{\rm conv} - \mathbb{P}^{\rm out}
\approx &     \frac{2}{D_{\rm W}} \int^{\frac{D_{\rm W}}{2}}_{0} \left[e^{-\phi  \sqrt{y^2+d^2} }\right. \\\nonumber &-\left. \frac{2}{D_{\rm L}} \int^{\frac{D_{\rm L}}{2}}_{0} e^{-\phi  \sqrt{x^2+y^2+d^2} } dx \right] d y.
\end{align}

Recall that for any $c>0$, the following inequality holds
 \begin{align}\label{convdfd}
   \frac{2}{D_{\rm L}} \int^{\frac{D_{\rm L}}{2}}_{0} e^{-\phi  \sqrt{x^2+c} } dx<    \frac{2}{D_{\rm L}} \int^{\frac{D_{\rm L}}{2}}_{0} e^{-\phi  \sqrt{c} } dx = e^{-\phi  \sqrt{c} } ,
   \end{align}
   which leads to the following corollary.
   \begin{corollary}\label{corollary1}
For the single-user special case with the blockage model in \eqref{blockage}, at high SNR, the outage probability achieved by pinching antennas is strictly smaller than that of conventional antennas.
\end{corollary}

Furthermore, we note that the performance gain $ \mathbb{P}^{\rm out}_{\rm conv} - \mathbb{P}^{\rm out}$ can be expressed as follows:
 \begin{align}\label{conv gap biger}
  \mathbb{P}^{\rm out}_{\rm conv} - \mathbb{P}^{\rm out} 
\approx &    \frac{2}{D_{\rm W}} \int^{\frac{D_{\rm W}}{2}}_{0} \left[e^{-\phi  \sqrt{y^2+d^2} }\right. \\\nonumber &-\left. 2\int^{\frac{1}{2}}_{0} e^{-\phi  \sqrt{D_{\rm L}^2z^2+y^2+d^2} } dz \right] d y,
\end{align}   
where $z=\frac{x}{D_{\rm L}}$. An important observation from \eqref{conv gap biger} is that   only the integral term, $\int^{\frac{1}{2}}_{0} e^{-\phi  \sqrt{D_{\rm L}^2z^2+y^2+d^2} } dx $, is a function of $D_{\rm L}$. Since increasing $D_{\rm L}$ reduces the value of this integral term, we can conclude that $  \mathbb{P}^{\rm out}_{\rm conv} - \mathbb{P}^{\rm out} $ is a monotonically increasing function of $D_{\rm L}$. 
    
\subsection*{Pinching Antennas in  Ultra-Dense Networks}
As shown previously, one of the key challenges to analyzing the outage probability is due to the square root function used by the LoS blockage model in \eqref{blockage}.  Recall that in ultra-dense indoor networks, the LoS blockage can be modeled differently, as shown in \eqref{blockage2}, where the squared distance is used in the exponent. By using this new model, the outage probability achieved by pinching antennas can be simplified as follows:
  \begin{align}\label{all terms2}
& \mathbb{P}^{\rm out} =  \frac{1}{D_{\rm W}}\int^{\frac{D_{\rm W}}{2}}_{-\frac{D_{\rm W}}{2}}\left(1-e^{-\phi   ({y^2+d^2}) }\right) d y  \\\nonumber &+  \frac{1}{D_{\rm W}}\int^{\tau_2}_{-\frac{D_{\rm W}}{2}}   e^{-\phi  ({y^2+d^2} ) }  d y   +  \frac{1}{D_{\rm W}}\int_{\tau_3}^{\frac{D_{\rm W}}{2}}   e^{-\phi  ({y^2+d^2)}  }  d y  .
\end{align}
The above expression can be evaluated without those high SNR approximations used previously. In particular, the outage probability can be first rewritten as follows:
  \begin{align}\label{all terms2}
& \mathbb{P}^{\rm out} =1- \frac{2}{D_{\rm W}}e^{-\phi    d^2 }  \int^{\frac{D_{\rm W}}{2}}_{0}e^{-\phi   y^2 } d y  \\\nonumber &+  \frac{1}{D_{\rm W}}e^{-\phi d^2 }   \int^{\tau_2}_{-\frac{D_{\rm W}}{2}}   e^{-\phi  y^2 }  d y   +  \frac{1}{D_{\rm W}}e^{-\phi  d^2  }\int_{\tau_3}^{\frac{D_{\rm W}}{2}}   e^{-\phi  y^2 }  d y  .
\end{align}
With some straightforward algebraic manipulations, a closed-form expression of $ \mathbb{P}^{\rm out}$ can be obtained, as shown in the following lemma.
\begin{lemma}\label{lemma2}
For the single-user special case with the blockage model in \eqref{blockage2}, the outage probability achieved by pinching-antenna systems can be expressed as follows:
 \begin{align}\label{all terms2}
& \mathbb{P}^{\rm out} = 1 -   e^{-\phi d^2 }  \frac{\sqrt{\pi}}{2\sqrt{\phi}D_{\rm W}}\left( \Phi\left(
-\sqrt{\phi}\tau_2
\right) +\Phi\left(
\sqrt{\phi}\tau_3
\right)\right),
\end{align}
where $\Phi(\cdot)$ denotes the error function \cite{GRADSHTEYN}.
\end{lemma}
 
The use of Lemma \ref{lemma2} yields the following high SNR approximation of $ \mathbb{P}^{\rm out}$:
 \begin{align}\label{blockage2ss}
\mathbb{P}^{\rm out} \approx  1-   \frac{\sqrt{\pi} e^{-\phi    d^2 }}{\sqrt{\phi}D_{\rm W}}\Phi\left(
\frac{\sqrt{\phi}D_{\rm W}}{2}
\right) .
\end{align}
Based on the blockage model in \eqref{blockage2}, the outage probability achieved by the conventional-antenna system can be approximated at high SNR as follows:
 \begin{align}\label{conv2}
 \mathbb{P}^{\rm out}_{\rm conv} \approx &\int^{\frac{D_{\rm L}}{2}}_{-\frac{D_{\rm L}}{2}}\int^{\frac{D_{\rm W}}{2}}_{-\frac{D_{\rm W}}{2}}\left(1-e^{-\phi   ({x^2+y^2+d^2}) }\right) \frac{d ydx   }{D_{\rm W}D_{\rm L}} 
\\\nonumber
=&1- \frac{  \pi e^{-\phi  d^2 }  }{D_{\rm W}D_{\rm L}\phi }  \Phi
\left(
\sqrt{\phi}\frac{D_{\rm L}}{2}
\right) \Phi
\left(
\sqrt{\phi}\frac{D_{\rm W}}{2}
\right) .
\end{align}
The difference between the outage probabilities achieved by the two considered schemes is given by 
 \begin{align}\label{conv2ddfd}
 \mathbb{P}^{\rm out}_{\rm conv} -\mathbb{P}^{\rm out} \approx &\gamma_1
 \left(1- \frac{  \sqrt{\pi}   }{ D_{\rm L}\sqrt{\phi} }  \Phi
\left(
\sqrt{\phi}\frac{D_{\rm L}}{2}
\right) \right) ,
\end{align}
where $\gamma_1= \frac{\sqrt{\pi} e^{-\phi    d^2 }}{\sqrt{\phi}D_{\rm W}}\Phi\left(
\frac{\sqrt{\phi}D_{\rm W}}{2}
\right)  $. 
Recall that the following inequality holds:
\begin{align}\nonumber
 \frac{ \sqrt{ \pi}    }{ D_{\rm L}\sqrt{\phi} }  \Phi
\left(
\sqrt{\phi}\frac{D_{\rm L}}{2}
\right)  =  \frac{ 2   }{ D_{\rm L}  }  \int^{ \frac{ D_{\rm L}  }{ 2   } }_{0} e^{-\phi x^2}dx<1. 
\end{align}
The above inequality leads to the conclusion that the outage probability achieved by conventional-antenna systems is strictly larger than that of pinching-antenna systems, which is consistent with Corollary \ref{corollary1}. Furthermore, the new result in \eqref{conv2ddfd} clearly reveals the impact of $D_{\rm L}$ on the performance comparison between the two considered schemes. Recall from \cite{GRADSHTEYN} that $\int^{\infty}_{0}e^{-q^2x^2}dx=\frac{\sqrt{\pi}}{2q}$, which leads to $\frac{ \sqrt{ \pi}    }{ D_{\rm L}\sqrt{\phi} }  \Phi
\left(
\sqrt{\phi}\frac{D_{\rm L}}{2}
\right)  =  \frac{ 2   }{ D_{\rm L}  }  \int^{ \frac{ D_{\rm L}  }{ 2   } }_{0} e^{-\phi x^2}dx\rightarrow  \frac{\sqrt{\pi}}{ D_{\rm L} \sqrt{\phi}}$, for $D_{\rm L}\rightarrow \infty$,   and hence $ \mathbb{P}^{\rm out}_{\rm conv} -\mathbb{P}^{\rm out} $ can be approximated as follows:
 \begin{align}\nonumber
 \mathbb{P}^{\rm out}_{\rm conv} -\mathbb{P}^{\rm out} \approx & \frac{\sqrt{\pi} e^{-\phi    d^2 }}{\sqrt{\phi}D_{\rm W}}\Phi\left(
\frac{\sqrt{\phi}D_{\rm W}}{2}
\right)     \left(1- \frac{  \sqrt{\pi}   }{ D_{\rm L}\sqrt{\phi} }   \right) ,
\end{align}
which confirms the previous conclusion that the performance gain of pinching antennas over conventional antennas is a monotonically increasing function of $D_{\rm L}$.
\section{Performance Analysis for the Case of $M>1$ }\label{section 4}
For multi-user cases, i.e., $M>1$, co-channel interference is a key factor limiting system performance. The aim of this section is to reveal that in pinching-antenna systems, LoS blockage is useful for suppressing co-channel interference. In order to illustrate this effect clearly, the ergodic data rate is used as the performance metric, and Design II is used as an example. Recall that the antenna spacing for conventional-antenna systems is half a wavelength, which means that the channels from different antennas to the same user are subject to the same blockage. Therefore,  for conventional-antenna systems,   ${\rm U}_m$'s ergodic data rate can be expressed as follows:
  \begin{align} \nonumber 
\bar{R}^{\rm Conv}_m =& \mathcal{E}\left\{ \log_2\left(
1+ \frac{ \alpha_{m}\left|h_{mm} \right|^2P}
{\alpha_{m}\sum_{i\neq m} \left|
h_{mi}   
\right|^2 P+M\sigma^2}
\right)
\right\}\\\label{conv bound} \approx &  \mathcal{E}\left\{ \log_2\left(
1+ \frac{  \left|h_{mm} \right|^2}
{ \sum_{i\neq m} \left|
h_{mi}   
\right|^2 }
\right)
\right\},
  \end{align}
where $\alpha_m$ denotes ${\rm U}_m$'s LoS blockage parameter, and the last step follows from a high SNR approximation. \eqref{conv bound} clearly shows that the ergodic data rate in conventional-antenna systems is bounded, and cannot be increased by increasing the transmit power only.

To facilitate the performance analysis, the case with $M=2$ is focused on in the following. Without loss of generality, we focus on ${\rm U}_1$'s data rate which can be simplified as follows:
\begin{align}\label{1pa1wg2}
  R^{\rm Pin}_1 = \log_2\left(
1+ \frac{\frac{\alpha_{11}\eta P}{ |\boldsymbol{\psi}^{\rm Pin}_1 - \boldsymbol{\psi}_1| ^{ 2 } }}{  \frac{\alpha_{12}\eta P}{ |\boldsymbol{\psi}^{\rm Pin}_2 - \boldsymbol{\psi}_1| ^{ 2 }}+M \sigma^2}
\right).
\end{align}
Hence, ${\rm U}_1$'s ergodic data rate can be expressed as follows: 
\begin{align}
  \bar{R}^{\rm Pin}_1  
= &\mathcal{E}\left\{\log_2\left(
1+ \frac{\frac{\alpha_{11}\eta P}{ (y_1-\beta_1)^2+d^2 }}{  \frac{\alpha_{12}\eta P}{(x_1-x_2)^2+(y_1-\beta_2)^2+d^2}+M \sigma^2}
\right)\right\}. 
\end{align}

By using the fact that the blockage coefficients, $\alpha_{mn}$, are discrete random variables, the ergodic data rate can be expressed as follows:
{\small \begin{align}\nonumber
  &\bar{R}^{\rm Pin}_1 = \mathcal{E}\left\{\mathbb{P}(\alpha_{11}=1,\alpha_{12}=0)\log_2\left(
1+ \frac{\frac{\alpha_{11}\eta P}{ |\boldsymbol{\psi}^{\rm Pin}_1 - \boldsymbol{\psi}_1| ^{ 2 } }}{  M \sigma^2}
\right)\right\}+ \\\nonumber
&\mathcal{E}\left\{\mathbb{P}(\alpha_{11}=1,\alpha_{12}=1)\log_2\left(
1+ \frac{\frac{\alpha_{11}\eta P}{ |\boldsymbol{\psi}^{\rm Pin}_1 - \boldsymbol{\psi}_1| ^{ 2 } }}{  \frac{\alpha_{12}\eta P}{ |\boldsymbol{\psi}^{\rm Pin}_2 - \boldsymbol{\psi}_1| ^{ 2 }}+M \sigma^2}
\right)\right\} .
\end{align}}
To facilitate the performance analysis, the blockage model in \eqref{blockage2} is used, and hence, the ergodic data rate is given by 
  \begin{align}  \label{complic} 
  \bar{R}^{\rm Pin}_1 =& \int^{\frac{D_{\rm L}}{2}}_{-\frac{D_{\rm L}}{2}}\int^{\frac{D_{\rm L}}{2}}_{-\frac{D_{\rm L}}{2}} \int^{\beta_1+\frac{D_{\rm W}}{2M}}_{\beta_0}
  \log_2\left(
1+ \frac{\frac{ \eta P}{g_1(y_1)  }}{   M \sigma^2}
\right) 
\\\nonumber &\times e^{-\phi  {g_1(y_1)  } }\left(1- e^{-\phi  {g_2(x_1,x_2,y_1) }}\right) \frac{dy_1dx_1dx_2}{D_{\rm L}^2D_{\rm W}} \\\nonumber
&+ \int^{\frac{D_{\rm L}}{2}}_{-\frac{D_{\rm L}}{2}}\int^{\frac{D_{\rm L}}{2}}_{-\frac{D_{\rm L}}{2}} \int^{\beta_1+\frac{D_{\rm W}}{2M}}_{\beta_0}e^{-\phi \left( {g_1(y_1)  }+ {g_2(x_1,x_2,y_1) }\right) }  \\ \nonumber
&\times \log_2\left(
1+ \frac{\frac{ \eta P}{ g_1(y_1) }}{  \frac{ \eta P}{g_2(x_1,x_2,y_1) }+M \sigma^2}
\right) \frac{dy_1dx_1dx_2}{D_{\rm L}^2D_{\rm W}},
\end{align}
where $g_1(y_1) =  (y_1-\beta_1)^2+d^2$, $g_2(x_1,x_2,y_1) = (x_1-x_2)^2+(y_1-\beta_2)^2+d^2$, and $\beta_0=-\frac{D_{\rm W}}{2}$.

The expression in \eqref{complic} contains multiple complicated integrals and hence is difficult to evaluate. In order to obtain insight, we focus on a special case with $y_m=\beta_m$, i.e., each user is right underneath its associated waveguide. In practice, this assumption can be justified for Internet of Things (IoT) applications for intelligent transportation and mining. With this assumption,  $ \bar{R}^{\rm Pin}_1$ can be approximated at high SNR as follows: 
 \begin{align} 
 \bar{R}^{\rm Pin}_1 \approx&  \int^{\frac{D_{\rm L}}{2}}_{-\frac{D_{\rm L}}{2}}\int^{\frac{D_{\rm L}}{2}}_{-\frac{D_{\rm L}}{2}}  \left(1- e^{-\phi {((x_1-x_2)^2+  \tau_4)}}\right)  \frac{dx_1dx_2}{D_{\rm L}^2}  \\\nonumber
& \times  \log_2\left(
1+ \frac{ \eta P}{ M \sigma^2  d^2 }
\right)
e^{-\phi d ^2} ,
\end{align}
where $\tau_4=(\beta_1-\beta_2)^2+d^2$.  
By using the fact that $x_1$ and $x_2$ are independently and uniformly distributed, the difference between the two random variables follows the triangular distribution, i.e.,
\begin{align}
f_{x_1-x_2}(z) =  \frac{D_{\rm L}-|z|}{D_{\rm L}^2},
\end{align}
for $-D_{\rm L}\leq z\leq D_{\rm L}$. 
Therefore, the ergodic data rate can be expressed at high SNR as follows:
\begin{align} 
  \bar{R}^{\rm Pin}_1 \approx&  \log_2\left(
1+ \frac{ \eta P}{ M \sigma^2 d^2 }
\right)
e^{-\phi d^2}
\\\nonumber &\times \int^{D_{\rm L}}_{-D_{\rm L}}  \left(1- e^{-\phi {(z^2+  \tau_4)}}\right)    \frac{D_{\rm L}-|z|}{D_{\rm L}^2}dz
\\\nonumber
=& 2 \log_2\left(
1+ \frac{ \eta P}{ M \sigma^2 d^2}
\right)
e^{-\phi d^2 }
\\\nonumber &\times \left(\frac{1}{2}-\int^{D_{\rm L}}_{0}  e^{-\phi  {(z^2+  \tau_4)}}\frac{D_{\rm L}-z}{D_{\rm L}^2}  dz\right)  .
\end{align}

With some straightforward algebraic manipulations, a high SNR approximation for ${\rm U}_1$'s ergodic data rate can be obtained as in the following lemma. 
\begin{lemma}\label{lemma3}
For the special case with $M=2$, assume that $y_m=\beta_m$ and the LoS blockage model in \eqref{blockage2} is used. ${\rm U}_1$'s ergodic data rate can be approximated at high SNR as follows:
\begin{align} \label{lemma3eq}
  \bar{R}^{\rm Pin}_1 \approx&   \left.  2 \log_2\left(
1+ \frac{ \eta P}{ M \sigma^2 d^2 }
\right)
e^{-\phi d^2 }\right[\frac{1}{2}- \frac{e^{-\phi   \tau_4} }{D_{\rm L}} 
\\\nonumber &  \left.\times \frac{\sqrt{\pi}}{2\sqrt{\phi}}\Phi\left(\sqrt{\phi} D_{\rm L}\right) + \frac{ e^{-\phi  \tau_4} }{2\phi D_{\rm L}^2}\left(
1-e^{-\phi  D_{\rm L}^2} 
\right)\right].
\end{align}
\end{lemma}

By using \eqref{conv bound} and \eqref{lemma3eq}, the following corollary can be obtained straightforwardly.
\begin{corollary}\label{corollary2}
For the two-user case, at high SNR, the performance gain of pinching antennas over conventional antennas is unbounded, i.e., $ \bar{R}^{\rm Pin}_1- \bar{R}^{\rm Conv}_1\rightarrow \infty$ for $\frac{P}{\sigma^2}\rightarrow \infty$.
\end{corollary}

The reason for this significant performance gain is that LoS blockage plays a positive role in pinching-antenna systems. In particular, because a user's distances to its interfering antennas are large, it is very likely that these interference links are subject to blockage, and hence co-channel interference can be effectively suppressed in pinching-antenna systems.

\section{Numerical Studies }
In the computer simulations, the noise power is $-90$ dBm,  $d=3$ m, $f_c=28$ GHz, $\lambda_{\rm cut}=10$ GHz, $\tilde{\Delta}=\frac{\lambda}{2}$ and $n_{\rm eff}=1.4$. Case I refers to the scenario with LoS blockage only, and Case II considers both LoS blockage and the waveguide propagation loss with $0.08$ dB/m \cite{mypa}.

\begin{figure}[t]\centering \vspace{-2em}
    \epsfig{file=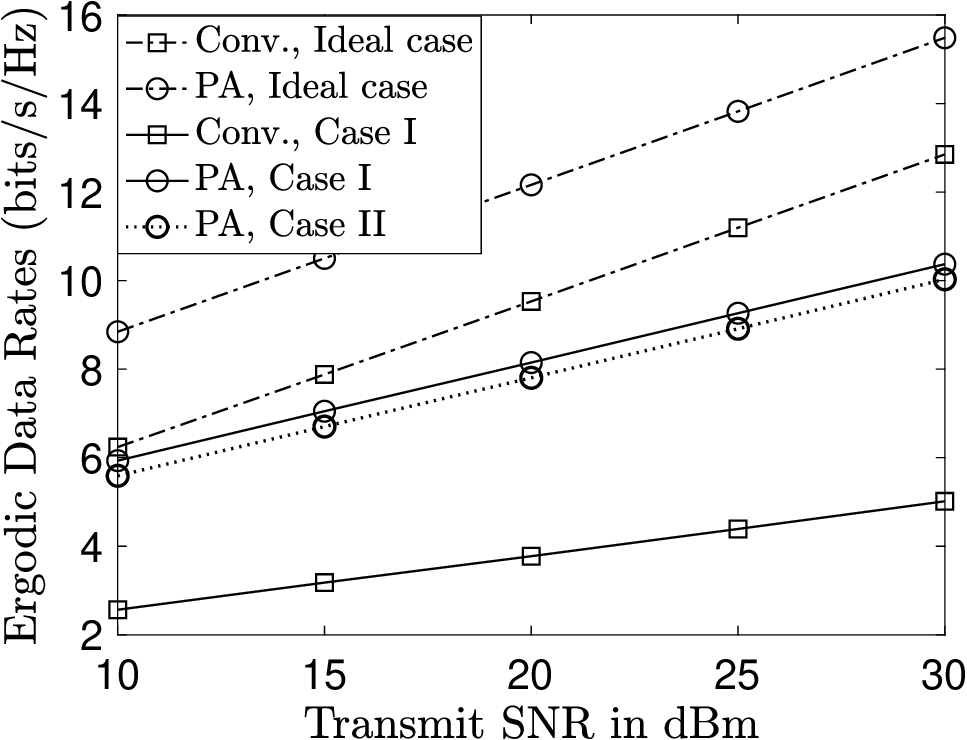, width=0.24\textwidth, clip=}\vspace{-0.5em}
\caption{{Ergodic data rates achieved by the considered transmission schemes for the single-user special case. $D_{\rm W}=10$ m,  $D_{\rm L}=4D_{\rm W}$, the blockage model in \eqref{blockage} is used, and $\phi=0.1$.   }
  \vspace{-1em}    }\label{fig1}   \vspace{-0.5em} 
\end{figure}

\begin{figure}[t] \vspace{-0.2em}
\begin{center}
\subfigure[Blockage Mode in \eqref{blockage}]{\label{fig2a}\includegraphics[width=0.24\textwidth]{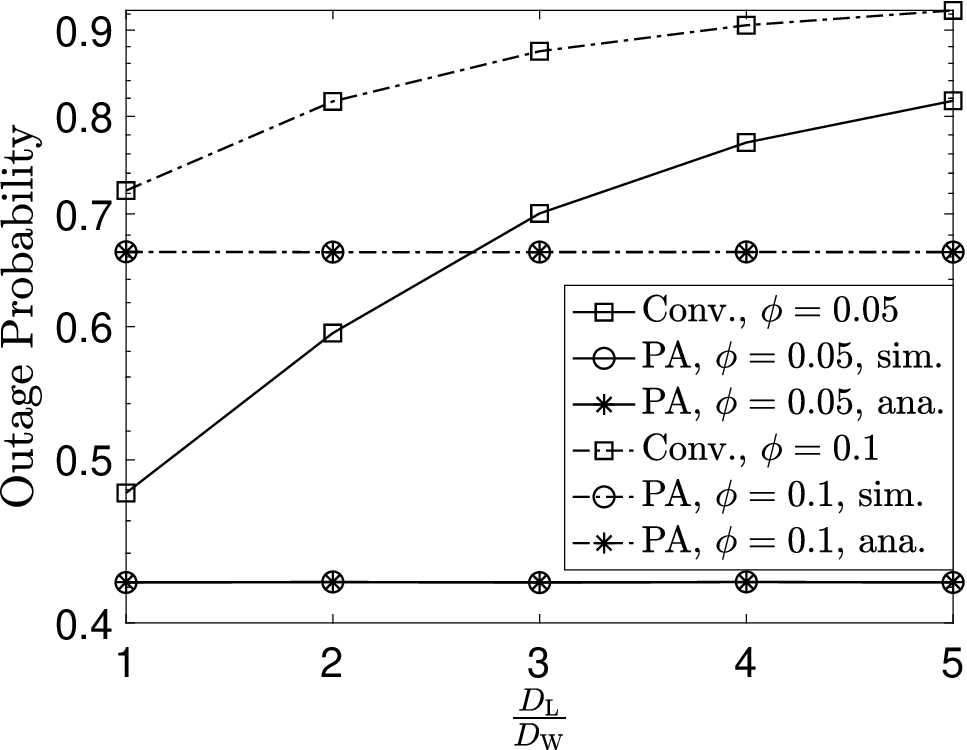}} 
\subfigure[Blockage Mode in \eqref{blockage2}]{\label{fig2b}\includegraphics[width=0.24\textwidth]{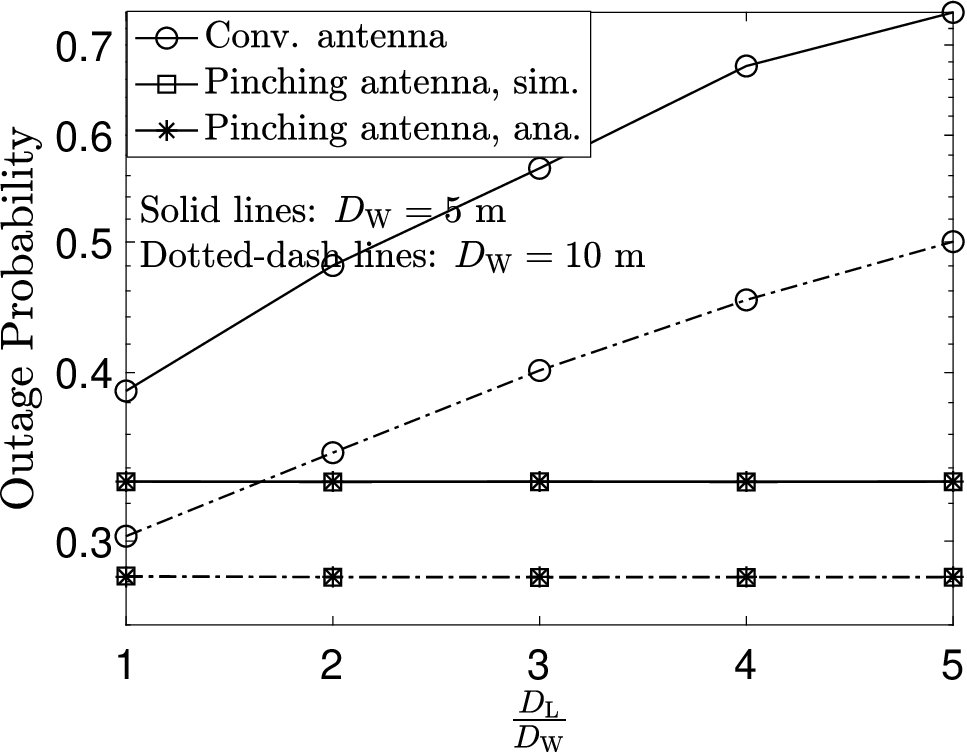}} \vspace{-1.5em}
\end{center}
\caption{Outage probabilities achieved by the considered transmission schemes for the single-user special case. In both figures, the transmit power is $10$ dBm, and Case II is considered. In  Fig. \ref{fig2a}, $\phi=0.1$. In Fig. \ref{fig2b}, $D_{\rm W}=5$ m, and the analytical results are based on Lemmas \ref{lemma1} and \ref{lemma2}.  \vspace{-1em} }\label{fig2}\vspace{-1.2em}
\end{figure}

Fig. \ref{fig1} shows the ergodic data rates achieved by the considered schemes. As can be seen from the figure, the pinching-antenna system is more robust to LoS blockage than the conventional one. Hence, in the presence of LoS blockage, the performance gain of pinching antennas over conventional antennas is much larger than that of the ideal case considered in \cite{mypa}. Fig. \ref{fig1} also shows that the effect of waveguide propagation loss is insignificant, which is consistent with \cite{mypa}.  In Fig. \ref{fig2}, the outage probability is used as the performance metric, where different blockage models are used. Fig. \ref{fig2} shows that, regardless of the used blockage model, the performance gain of pinching-antenna systems over conventional-antenna systems is significant, particularly with large $D_{\rm L}$.  
  
   \begin{figure}[t] \vspace{-2em}
\begin{center}
\subfigure[$M=2$]{\label{fig3a}\includegraphics[width=0.24\textwidth]{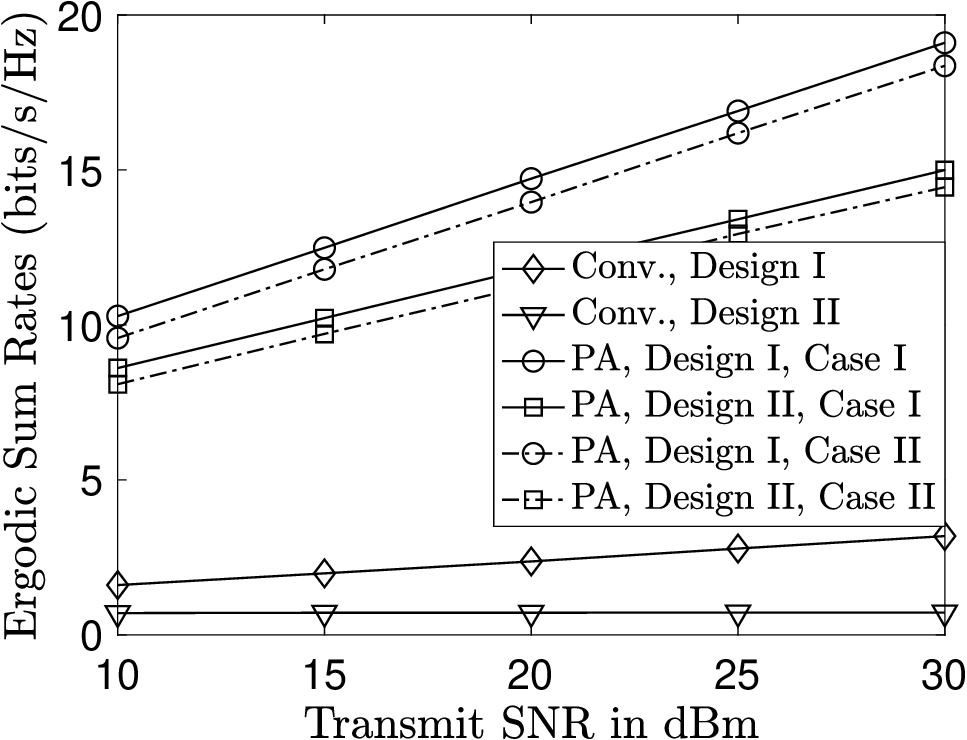}} 
\subfigure[$M=5$]{\label{fig3b}\includegraphics[width=0.24\textwidth]{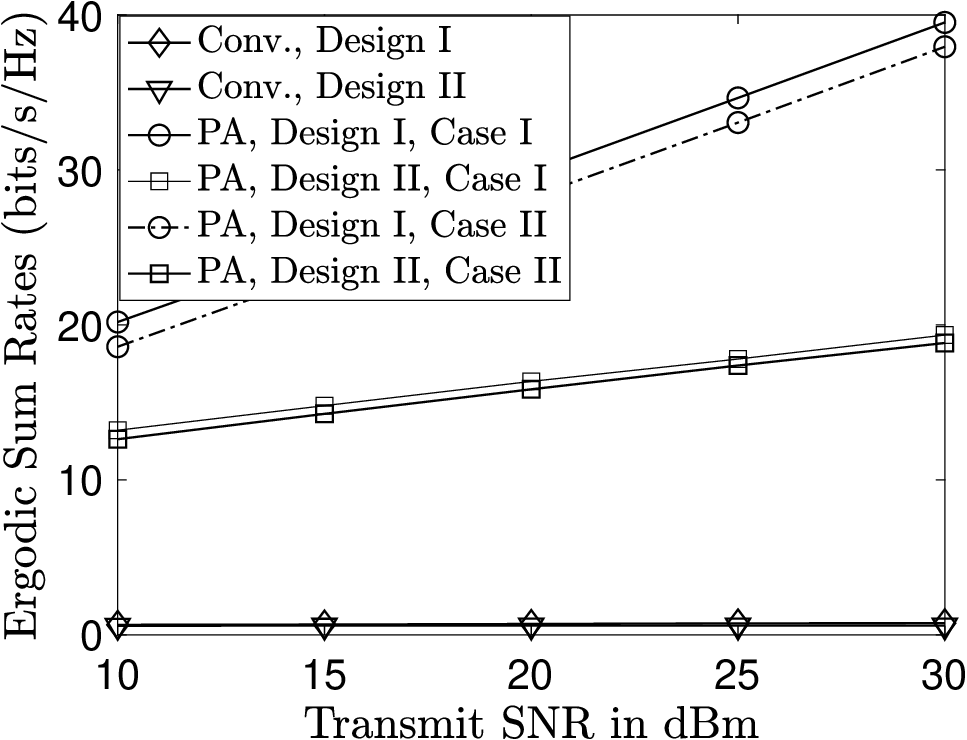}} \vspace{-1.5em}
\end{center}
\caption{Ergodic sum rates achieved by the considered transmission schemes for the multi-user case. In both figures, $D_{\rm W}=10$ m, $D_{\rm L}=4D_{\rm W}$, the blockage model in \eqref{blockage} is used, and $\phi=0.1$.     \vspace{-1em} }\label{fig3}\vspace{-0.2em}
\end{figure}

    \begin{figure}[t]\centering \vspace{-0em}
    \epsfig{file=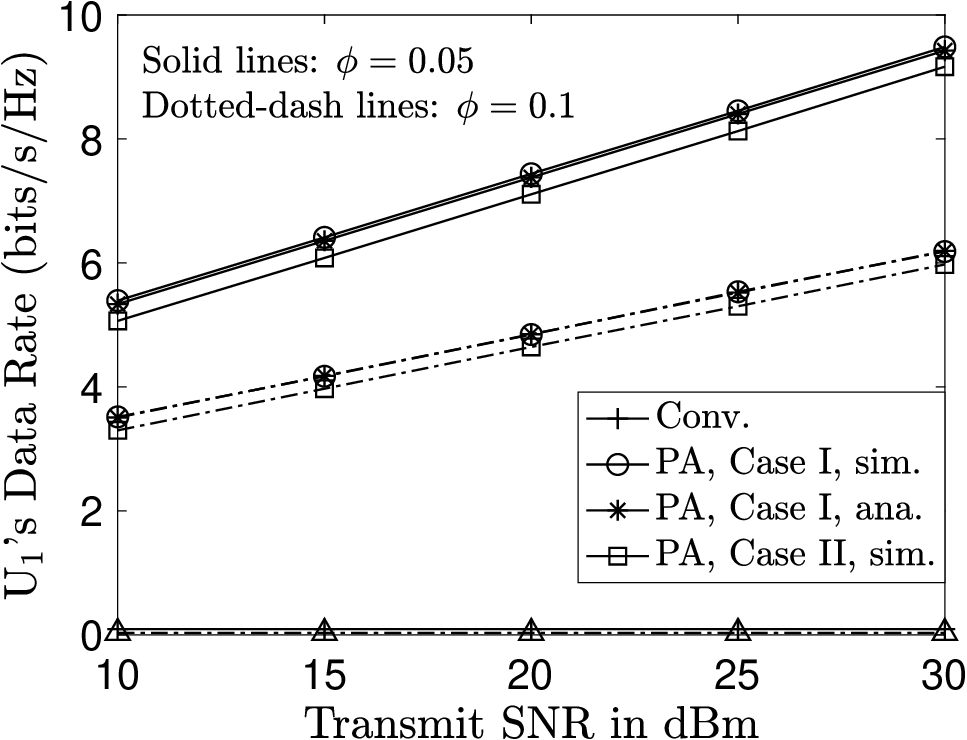, width=0.24\textwidth, clip=}\vspace{-0.5em}
\caption{{ Ergodic data rates achieved by the considered transmission schemes for the special case considered in Section \ref{section 4}.   $D_{\rm W}=10$ m,  $D_{\rm L}=4D_{\rm W}$, Design 2 is considered, and the blockage model in \eqref{blockage2} is used. The shown analytical results are based on Lemma \ref{lemma3}. }
  \vspace{-1em}    }\label{fig4}   \vspace{-0.5em} 
\end{figure}


Fig. \ref{fig3} focuses on the general multi-user case, by using the ergodic sum rate as the performance metric. Fig. \ref{fig3} demonstrates that the performance gain of pinching-antenna systems over conventional ones in the multi-user case becomes much more significant than that of the single-user case. Fig. \ref{fig3} also shows that the performance gain of pinching-antenna systems is a monotonically increasing function of the SNR, which confirms Corollary \ref{corollary2}. Furthermore, the low-complexity design, Design II, might lead to a performance loss to Design I; however,  there is still a significant performance gain over conventional antenna systems. In Fig. \ref{fig4}, the blockage model in \eqref{blockage2} is used. As consistent with Fig. \ref{fig3}, Fig. \ref{fig4} shows that the performance gain of pinching-antenna systems over conventional-antenna systems is significant. In addition, Figs. \ref{fig2} and \ref{fig4} also show that the curves of the analytical results perfectly match those simulation-based ones, which verifies the accuracy of the developed analytical results. 
  \vspace{-0.5em}
\section{Conclusions}
This letter has investigated the impact of LoS blockage on pinching-antenna systems. Analytical results have been developed for both the single-user and multi-user cases to reveal that the presence of LoS blockage is beneficial for increasing the performance gain of pinching antennas over conventional antennas. LoS blockage is particularly useful in multi-user cases, where co-channel interference can be effectively suppressed.   
  \vspace{-0.5em}
\bibliographystyle{IEEEtran}
\bibliography{IEEEfull,trasfer}
  \end{document}